\documentclass[twocolumn]{emulateapj}
\newcommand{\kms}{\rm km~s^{-1}}

\slugcomment{Submitted to the Astrophysical Journal Letters}

\shorttitle{Bulge Radial Velocity Assay}
\shortauthors{Rich et al.}

\begin{document}

\title{The Bulge Radial Velocity Assay (BRAVA): I. Techniques
and a Rotation Curve}

\author{R. Michael Rich\altaffilmark{1}, David B. Reitzel\altaffilmark{1,3}, and Christian D. Howard\altaffilmark{1}}

\author{HongSheng Zhao\altaffilmark{2}}

\altaffiltext{1}{Department of Physics and Astronomy, UCLA, Los Angeles, CA 90095-1547 rmr@astro.ucla.edu}
\altaffiltext{2}{SUPA, School of Physics and Astronomy, University of St Andrews, KY16 9SS, UK}
\altaffiltext{3}{Visiting Astronomer, Cerro Tololo Inter-American Observatory.
CTIO is operated by AURA, Inc.\ under contract to the National Science
Foundation.}

\begin{abstract}

We are undertaking a large scale radial velocity survey of the
Galactic bulge which uses M giant stars selected from the {\sl 2MASS}
catalog as targets for the CTIO 4m Hydra multi-object spectrograph.
The aim of this survey is to test dynamical models of the bulge and to
quantify the importance, if any, of cold stellar streams in the bulge
and its vicinity.  Here we report on the kinematics of a strip of
fields at $-10^{\circ} < l <+10^{\circ}$ and $b=-4^{\circ}$.  We construct a
longitude-velocity plot for the bulge stars and the model data, and find
that contrary to previous studies, the bulge does not
rotate as a solid body.   From $-5^{\circ}<l<+5^{\circ}$ the rotation curve 
has a slope of roughly
$100~\rm \kms~kpc^{-1}$  and flattens considerably
at greater $l$ and reaches a maximum rotation
of $45~\kms$.   We compare our rotation curve
and velocity dispersion profile to both the self-consistent
model of \citep{zhao96} and to N-body models; neither
fits both our observed rotation curve
and velocity dispersion profile.  The high precision of our radial
velocities $(\sim 3~\kms )$ yields an unexpected result: hints of cold
kinematic features are seen in a number of the line of sight velocity
distributions.

\end{abstract}


\section{Introduction}

The {\sl COBE} $2 \micron$ image of the bulge \citep{dwek95, binney97} 
and models of the projected 2$\mu \rm m$ light shows
a bar-like structure that is also detected in star counts of
red clump stars \citep{stanek97}.  The anomalously high optical depth
of the bulge to microlensing \citep{alcock00} can be explained only
by assuming a bar whose major axis extends roughly toward the Sun,
thus raising the rate of star-star events \citep{han03}.  Theoretical
models of the bulge initially followed axisymmetric models
\citep{kent92} but have graduated to self-consistent rapidly rotating
bars \citep{zhao96, hafner00,bissantz04}, with the density and the potential strongly 
constrained by the observed microlensing 
rates in the bulge as well as gas kinematics.  However, the phase space of the bar is
relatively incompletely constrained by stellar kinematic data.

Study of the kinematics of the bulge is complicated by the large and
variable foreground extinction, the presence of a contaminating disk
population extending from the foreground well into the Galactic
Center, and source confusion arising from the high density 
of stars.  Positional measurements from wide field Schmidt plates are
therefore impossible, where source confusion makes any astrometric
exercise daunting.

M giants, while faint in the traditional optical bandpasses due to their
cool temperatures and TiO bands, are bright in the $I$ band
$(13<I<11)$ and are easy targets for spectroscopy, if positions
are known.  They further have the advantage of being ubiquitous
throughout the bulge and are luminous enough to be studied even
in fields with substantial extinction.  Finally, the short lifetimes
of AGB stars and luminous giants limit their numbers enough that
source confusion is not an issue.

\citet{mould83} was the first to measure the velocity dispersion of
the bulge in Baade's Window, based on the M giant sample of \citet{bmb84}
that was first classified from low dispersion slitless spectra.
Multi-fiber spectroscopy of this sample \citep{scw90} subsequently 
harvested roughly 250 bulge giant velocities.  Despite numerous
investigations of the dynamics of stars in the direction of Baade's
Window \citep{mould83,rich90,scw90,srt96} and in other bulge fields
\citep{tyson91,min92,min96,blum94,blum95} there has been, up 
to now, no large scale survey of the dynamics of the
stellar population in the bulge. \citet{beau00} survey the planetary
nebula population; both the rarity of PNe (due to their brief
lifetimes), the problem of disk or bulge membership, and the
considerable distance uncertainty make PNe a problematic population;
We believe a well selected sample of M giants is likely the best probe of the
bulge/bar population.

\citet{fw87} showed that the M giant luminosity function is
consistent with an old population roughly the age of the halo and
globular clusters. The first detailed abundance analysis of M giants
\citep{rich05} finds an abundance range similar to that of the K
giants \citep{mr94,fmr06}.  In principle, the most metal-poor
component of the bulge population might not evolve through the M giant
phase as the K giant abundance distribution is peaked at slightly
sub-solar metallicities, but most of the stars are more metal-rich 
than 47 Tuc (at $-0.7$ dex) and are therefore candidates to
reach the luminosities and effective temperatures characteristic
of the M giants \citep{zoc03}.

The proper motion study of \citet{sumi04} addresses a number of fields 
in the bulge (avoiding high extinction) and eventually, a second epoch
of {\sl 2MASS} imaging would provide proper motions for large numbers of
giants.  However, the addition of radial velocities and ultimately
metal abundances, is needed for a complete dynamical model and the M
giant population provides the perfect sample of stars to target.

\begin{figure}[t]
  \epsscale{1.0}
  \plotone{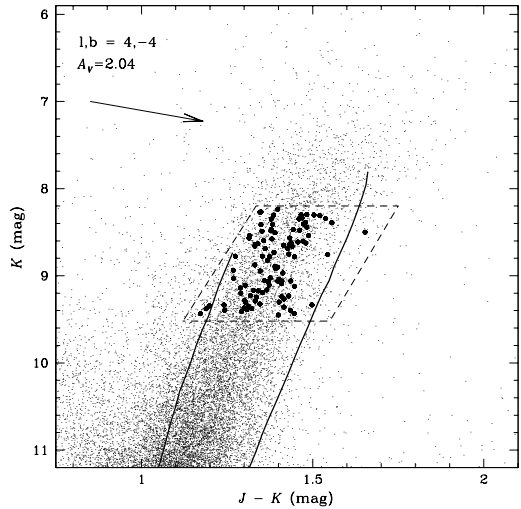}
  \caption{Fig. 1:   
 Color-magnitude diagram of {\sl 2MASS} candidates and
 filled symbols (observed stars), including reddened isochrones \citep{gir02}
 for [F/eH]=$-1.3$ and $-0.5$.  The parallelogram
 indicates our selection region; the  blue cutoff rejects many objects
 that are closer than the bulge, which  have lower reddening and are
 brighter than the red giant branch.  The reddening vector
 corresponds to $E(J-K)=0.33$ from the \citet{schlegel} map.  \label{fig1}} 
\end{figure}

The dynamical model for the bulge/bar has a number of important
implications.  Large samples of uniform radial velocity data are still
of great value in constraining the bar vs. axisymmetric models, and
the nature of the orbit families supporting the bar.  Further, the
interpretation of the microlensing events in the bulge depends on the
use of an accurate dynamical model \citep{han03}.  The recent
discovery of planetary transit host stars in the bulge \citep{sahu06}
gives an additional incentive to improve our
knowledge of the bulge/bar model.

With the availability of the Two Micron All-Sky Survey  {\sl 2MASS}
\citep{2mass} we realized that the key ingredients of high precision
positions and photometry would finally be available everywhere in the
Galactic bulge except near the plane of the Galaxy.  At this time, we
present $\sim 2300$ spectra and have obtained a radial velocity
precision of $\approx  3~\kms $ for our most recent (2006) data.  

In the past, optical radial velocity studies in the bulge have not emphasized
high precision, because of the large velocity dispersion and the
expectation that the short orbital periods would phase-wrap any cold
structures out of existence in well under a Gyr.  Yet our precision is
sufficient to enable a search for more cold streams analagous to those
associated with the Sagittarious dwarf spheroidal galaxy; some candidate
cold features are seen and followup observations are underway (Reitzel et al. 2007 in prep.).  Here we report on the kinematics of stars along a band at $b=-4^{\circ}$, obtaining a rotation curve and velocity dispersion profile.  

\section{Observations and Sample Selection}

The choice for optical spectroscopy is driven by
the availability of multi-object wide field spectroscopy.
IR spectroscopy, such as that of \citet{blum95}, can
be used to great effect in the fields of highest reddening
but obviously yields far fewer spectra.  Here we present
a brief sketch of the selection method and analysis; full
details will be given in Howard et al.(2007 in prep).
   
We use the survey method of \citet{scw90} as the model for our
program.  They observe that stars with $I<11.8$ have a lower velocity
dispersion and are likely disk members.  When the same field is
examined in {\sl 2MASS}, the $K$ vs $J-K$ CMD shows a clearly defined red
giant branch.  In the SCW90 field, we find that the magnitude limit of
$I<11.8$ corresponds roughly to $K<8$.  Further, there is break in the
luminosity function at this $K$ magnitude \citep{fw87}.  While a few
bulge AGB members may be present at brighter magnitudes, the bulk of
such bright stars will be foreground contaminants.  We adopt a range
of $9.25<K<8$ as the basis for selecting the {\sl 2MASS} M giants (Figure 1).

A parallelogram-shaped selection region is adjusted in color and
magnitude to encompass completely the red giant branch locus
of old stars at the distance and reddening of the field as indicated by the
reddening map of \citet{schlegel}.  We find that reddening varies greatly,
ranging from $1.5<A_V<5.0$.  In fields with high extinction,
differential reddening is an issue and we widen the selection region
to accommodate the red giant branch as observed (Figure 1).  The indicated
isochrones show that we admit stars with  $-1.3<$[Fe/H]$<0.3$ a range
that spans the entirety of the \citet{zoc03} abundance distribution

\subsection{Spectroscopy}
                        
\begin{figure}[t]
%\plotone{rich_fig2.eps}
%\epsscale{0.9}
\includegraphics[angle=-90,scale=.34]{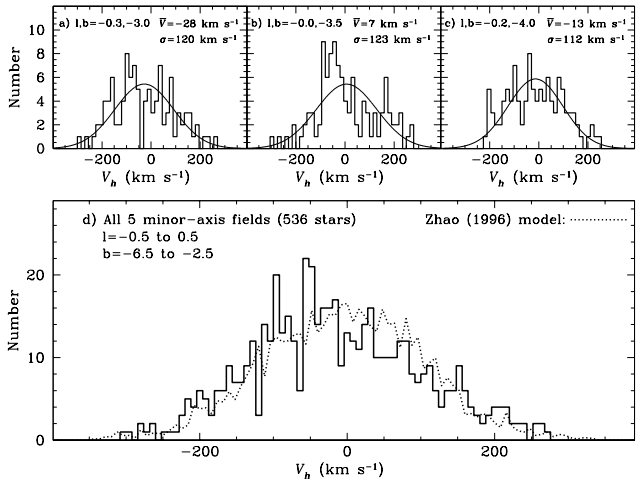}
\caption{Fig. 2a-c.: Velocity dispersion profile of three minor axis fields
  ranging from $-3^{\circ}<b<-4^{\circ}$; note the clumpiness in a and b.  
  The velocity dispersion profile of the sum of these fields is given
  in Fig 2d.  The dotted line is the prediction from the
  \citet{zhao96} model.  
  \label{fig2}} 
\end{figure} 
    
\begin{figure}[t]
\epsscale{0.9}
\plotone{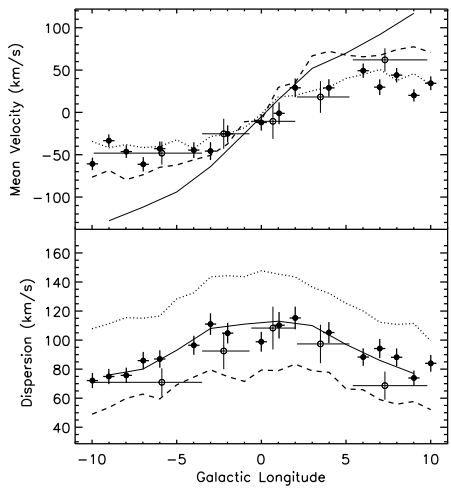}
\caption{Fig. 3a: Rotation curve from our data (filled symbols); open symbols
indicate binned PNe from \citet{beau00}; the PNe agree reasonably well with
our data and the departure from solid body rotation is clear.
The solid line indicates the model of \citet{zhao96} while the dotted
line is that of \citet{fux97} and the dashed line is from \citet{sell93} (see text).   Fig. 3b: Velocity dispersion
  is indicated with the PNe and models as above. 
  \label{fig3}}
\end{figure}

We use the Hydra multi-fiber spectrograph at the cassegrain focus of
the Blanco 4m telescope at Cerro Tololo.  We optimize for spectroscopy
in the red, taking advantage of the red colors of M giants and employ
the KPGLD grating, blazed at 8500\AA \ giving 0.45\AA $\ {\rm
pix}^{-1}$ with 2-pixel on-chip binning yielding an effective resolution of
0.88\AA \ ${\rm pix}^{-1}$ and a full spectra range of 1800\AA. We use
 the $200\mu \rm m$ slit plate, giving us an increase in
resolution; the loss of light from the slit plate is
inconsequential to our S/N because the stars are so bright.  Our
useful spectral range in 2006 was 6891\AA \ to 8714\AA;  the first two lines of the $\rm Ca {\sc II}$ infrared triplet are often well detected and 
contribute to our success in obtaining a correlation peak.  However,
there are also stars that are so red that the Ca triplet is overwhelmed
by TiO absorption; in these cases, the wide wavelength range is
essential for the velocity measurement.
 Our exposures are typically
$3\times 900~s$ but longer exposures were needed when cirrus was
present.  Each field has on average, 108 successfully exposed M
giants and 20 fibers that are used to obtain a sky background
spectrum.  Flat fielding, sky subtraction, throughput correction,
scattered light subtraction, and wavelength calibration for each
exposure were all accomplished using the IRAF task dohydra.
The spectra are binned to 34.3~$\kms~pix^{-1}$ and
normalized.  Radial velocities are measured using the fxcor task in
IRAF; this requires the spectra to be Fourier filtered exclude
features greater than 50 pixels or smaller than 3 pixels in extent.
Regions of the spectrum with obvious telluric features were excluded,
leaving only 60\% of the spectrum usable for cross correlation.  Our
final cross correlation regions are 7000-7150\AA, 7300-7580\AA,
7700-8100\AA, and 8300-8600\AA.

3 standard stars are all used in the cross correlation, with the final
velocity for any given star being the average of the 3 derived
velocities.  The three standards return velocities that agree to
within $1.6~\kms$ on average, with a standard deviation of
$1.4~\kms$ in these differences.

We report radial velocities for a total of 2294 M giants.
Here we consider the fields spanning across $b=-4^{\circ}$.  We
compare our observed velocities and velocity dispersion with those
predicted by the self consistent rotating bulge/bar model of
\citet{zhao96}.

\section{The Rotation Curve and Velocity Dispersion profile}

We now discuss the results from our study, beginning with the minor
axis velocity dispersion (Figure 2) resulting from the
sum of all fields on the minor axis at $b=-2.5^{\circ}$ to
$-6.5^{\circ}$ as well as three examples of the fields that went into
the summed distribution. The best fit Gaussian to the data
gives $\sigma = 119 \pm 5~\kms$ with $\rm v_0=-11 \pm 30~\kms$.  A
number of peaks are present in this histogram; these are prominent in
the contributing histograms (Figure 2a-c).  A comparison with the
equivalent region extracted from the \citet{zhao96} model is
also shown; the agreement is very good, but the velocities in the data
appear to be more clumpy than they are in the model.  More
striking peaks $(\approx 2.5\sigma)$  are are found in some of our other fields, but because such features can occur in random draws, new
observations are required for confirmation (Reitzel et al. 2007).

\begin{figure}[t]
\epsscale{0.9}
\plotone{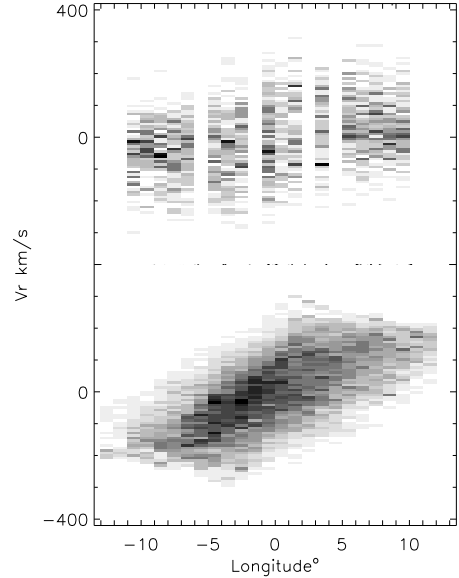}
\caption{Fig. 4:  
  {\it Upper panel:} longitude-velocity diagram for the data along
  $b=-4^{\circ}$, which have been smoothed to $1^{\circ}$ bins and
  by $8~\kms$.   
  {\it Lower Panel:}  The same region extracted from the
  \citet{zhao96} model.  The  lower right region would be
  populated by stochastic orbits in the model.  By
  construction, the model has no retrograde orbits; 
  their inclusion would help bring the model
  into agreement with the data.  
  \label{fig4}} 
\end{figure}

Figure 3a,b shows our rotation curve and velocity dispersion profile
compared with that predicted by \citet{zhao96}; we give the data in
Table 1.   We do not confirm the solid body rotation claimed by
a number of previous studies;
after reaching an amplitude of $\sim 40~\kms$, the
rotation curve flattens beyond $|{l}|>3^{\circ}$.  The velocity
dispersion profile remains higher than $75~\kms$ even for the fields
at $l=10^{\circ}$.  We compare our data with the planetary nebulae (PNe)from \citep{beau00}.  In order
to have a reasonable number of PNe, we have accepted those in the
range $-8^\circ <b<-3^\circ$ and we have binned the data.  Considering the less
secure distances and assignment of population for the PNe, the agreement
is good, and settles the question of solid body rotation for the bulge.  The
\citet{zhao96} model is a self-consistent rapidly rotating bar that is
constrained to have no retrograde orbits.  Also plotted are the N-body
bars of \citet{fux97} and \citet{sell93}, both of which are N-body
bars formed from initially unstable disks.  Sellwood's bar starts from
a rigid Plummer sphere with a live Kuzmin disk, with mass ratio 3:7 and
no dark matter.  Fux's bar starts from an equilibrium of dark halo, power
law nucleus, and an exponential disk; in contrast to the Zhao model that
is fit to the data, a best-fit model is selected from a large number of
N-body realizations.

We adopt the bar angles and mass normalisations as suggested
by \citet{beau00} that give an overall good match with
the appearance of the COBE map.
We project these models and calculate the velocity moments
in each line of sight without
distinguishing disk and bulge particles.
Neither model is satisfactory:
Fux's best model gives a remarkably good fit
to the rotation curve but has too low $V_{rot}/\sigma$.
Sellwood's bar appears to have a very high $V_{rot}/\sigma$.
It appears that our data is challenging to fit by
both disk-instability formed bars as well
as Schwarzschild models with fixed potentials.

In Figure 4, we compare the \citet{zhao96} model in longitude-velocity
space with our radial velocities, both sampling the slice at $b=-4^o$;
we have smoothed the data in longitude by $1.0^o$ and $8~\kms$ in
velocity.  The addition of retrograde orbits may improve the agreement
of the model and data; we are working toward this end (Zhao et al. 2007 in prep.).

While our claim of slower rotation for the bulge appears to contradict earlier studies, we emphasize our agreement with the PNe.   It is also interesting that the kinematics of K giants in the $(l,b)=(8,7)$ field of \citet{min92} agrees with our rotation
curve (44.5~$\kms$); faster rotation (and a lower velocity dispersion)
is observed in his $(l,b)=(12,3)$ but that field that might be disk dominated.
\citet{menz90} used a small number of Miras to sample velocities over a wide range in
longitude, and found rapid solid body rotation.
Small numbers, or contamination by members of the disk, might be invoked to explain the more rapid rotation observed for SiO masers \citep{izu95}.

Our data are clearly at odds with the widely held view that the bulge rotates
as a solid body.  The relatively high velocity dispersions of our
fields is reassuring in the sense that the M giant selection criterion
is yielding good kinematic probes of the bulge/bar population.  
Extending this survey to a larger number of fields in the bulge offers
the possibility of undertaking detailed tests of dynamical models,
something that is not presently possible in distant galaxy
populations.  This will give new insights into the structure and
dynamics of the bulge, and into the formation of the Milky Way.

\acknowledgments  

RMR, DBR, and CDH acknowledge support from STScI grant GO-10816 and
NSF grant AST-0709731.  We thank  the staff of CTIO, especially Knut Olsen and
Angel Guerra, and Andreas Koch and Jan Kleyna
for helpful comments on drafts.   We use data products from the
Two Micron All Sky Survey, which is a joint project of the University
of Massachusetts and the Infrared Processing and Analysis
Center/Caltech, funded by NASA and the NSF.
{\it Facilities:} \facility{CTIO}.

\clearpage
\begin{deluxetable}{ccccccccc}
\tabletypesize{\scriptsize}
\tablecaption{Observed Bulge Rotation Curve and Velocity Dispersion
Profile \label{tbl-1}}
\tablewidth{0pt}
\tablehead{
\colhead{l} & \colhead{b} & \colhead{RA (J2000.0)} & \colhead{DEC
J2000.0)} & \colhead{N} &
\colhead{$v$} & \colhead{err($v$)} & \colhead{$\sigma$} &
\colhead{err($\sigma$)} \\
\colhead{(degrees)} & \colhead{(degrees)} & \colhead{(hh:mm:ss.s)} &
\colhead{($^{\circ}$:$^{\prime}$:$^{\prime\prime}$)} & \colhead{} &
\colhead{($\kms$)} & \colhead{($\kms$)} & \colhead{($\kms$)} &
\colhead{($\kms$)}}

\startdata
-9.98 & -3.99 &  17:36:31.47 &  -39:30:42.9 &  93 & -60.8 &   7.5 &  72.1
&   5.3 \\
-9.01 & -4.00 &  17:39:10.83 &  -38:41:41.7 & 101 & -33.4 &   7.5 &  74.9
&   5.3 \\
-7.98 & -4.00 &  17:41:56.69 &  -37:49:13.2 & 101 & -46.3 &   7.5 &  75.7
&   5.3 \\
-6.98 & -3.99 &  17:44:33.22 &  -36:57:55.3 & 103 & -61.3 &   8.5 &  85.9
&   6.0 \\
-6.00 & -3.99 &  17:47:03.54 &  -36:07:46.5 & 105 & -42.8 &   8.5 &  87.0
&   6.0 \\
-3.99 & -3.98 &  17:52:02.74 &  -34:24:11.0 & 108 & -44.6 &   9.3 &  96.4
&   6.6 \\
-3.01 & -3.99 &  17:54:26.79 &  -33:33:26.6 & 111 & -45.7 &  10.5 & 111.0
&   7.5 \\
-1.99 & -3.98 &  17:56:50.87 &  -32:40:49.5 & 108 & -25.3 &  10.1 & 104.8
&   7.1 \\
-0.02 & -3.99 &  18:01:28.27 &  -30:58:18.6 & 110 & -14.7 &  10.8 & 112.8
&   7.6 \\
 1.04 & -3.95 &  18:03:43.51 &  -30:01:49.9 &  73 &  -1.1 &  12.9 & 110.2
&   9.1 \\
 2.00 & -4.00 &  18:06:02.64 &  -29:13:17.4 & 109 &  29.0 &  11.0 & 115.2
&   7.8 \\
 3.99 & -3.98 &  18:10:19.36 &  -27:28:32.3 & 106 &  29.0 &  10.2 & 105.2
&   7.2 \\
 6.01 & -3.99 &  18:14:40.05 &  -25:42:06.9 & 108 &  49.3 &   8.5 &  88.3
&   6.0 \\
 6.99 & -3.96 &  18:16:34.61 &  -24:49:41.9 & 104 &  29.8 &   9.2 &  94.2
&   6.5 \\
 7.99 & -3.99 &  18:18:44.88 &  -23:57:51.7 & 107 &  44.0 &   8.5 &  88.2
&   6.0 \\
 9.01 & -3.98 &  18:20:47.51 &  -23:03:31.1 & 104 &  20.1 &   7.2 &  73.9
&   5.1 \\
10.00 & -3.99 &  18:22:50.13 &  -22:11:12.6 & 106 &  34.4 &   8.2 &  84.0
&   5.8 \\

\enddata

\end{deluxetable}

\end{document}